\documentclass[prb,showpacs,a4paper,floatfix,final,twocolumn]{revtex4}
\usepackage{graphicx,epsf}
\usepackage{amssymb,amsfonts,amsmath}
\usepackage{times}
\usepackage{psfrag}
\usepackage{pst-all}
\usepackage{tabularx}

\newcommand{\Tcr}{T_\mathrm{c}}

\begin{document}
\title{Monte Carlo simulations of the directional-ordering transition in
  the two-dimensional classical and quantum compass model}
\date{\today}
\author{Sandro \surname{Wenzel}}
\email{wenzel@itp.uni-leipzig.de}
\author{Wolfhard \surname{Janke}}
\email{janke@itp.uni-leipzig.de}
\homepage{http://www.physik.uni-leipzig.de/cqt.html}
\affiliation{Institut f\"ur Theoretische Physik and Centre for Theoretical Sciences (NTZ), Universit\"at Leipzig, Postfach 100920, D-04009 Leipzig, Germany}

\begin{abstract}
  A comprehensive study of the two-dimensional (2D) compass model on the square lattice is
  performed for classical and quantum spin degrees of freedom using
  Monte Carlo and quantum Monte Carlo methods. We employ
  state-of-the-art implementations using Metropolis, stochastic series
  expansion and parallel tempering techniques to obtain the critical
  ordering temperatures and critical exponents. In a pre-investigation
  we reconsider the classical compass model where we study and
  contrast the finite-size scaling behavior of ordinary periodic
  boundary conditions against annealed boundary conditions.  It is
  shown that periodic boundary conditions suffer from extreme
  finite-size effects which might be caused by closed loop excitations
  on the torus. These excitations also appear to have severe effects on
  the Binder parameter. On this footing we report on a systematic
  Monte Carlo study of the quantum compass model. Our numerical
  results are at odds with recent literature on the subject which we
  trace back to neglecting the strong finite-size effects on periodic
  lattices. The critical temperatures are obtained as
  $T_\mathrm{c}=0.1464(2)J$ and $T_\mathrm{c}=0.055(1)J$ for the
  classical and quantum versions, respectively, and our data support a
  transition in the 2D Ising universality class for both cases.
\end{abstract}
\maketitle

\section{Introduction}

The compass model is one of the simplest models possessing orbital
degenerate states. Originally developed\cite{kugel82} as a model for
Mott insulators it has recently seen renewed
interest\cite{nussinov_orbitalorder,khomskii-2003,mostovoy:167201} in
connection with orbital-order in materials like transition metal (TM)
compounds. Despite its closeness to ordinary models of quantum
magnetism, like the Heisenberg model, there is no ordered phase
characterized by magnetization properties. This means that the ordered
phase appearing in the model is especially interesting in that it
cannot be classified according to the Mermin-Wagner criterion.
A competition
of interactions in different directions rather results in a special
long-range ordered state\cite{mishra:207201} possessing a sense of
orientation,\footnote{Hence the name compass model.} and the transition
is at the same time accompanied by dimensional
reduction.\cite{Batista_dimred2005} The current interest in this model
is furthermore triggered by the recent discovery that it
describes arrays of superconducting Josephson junctions and because of a
possible realization of a system which protects qubits against
unwanted decay in quantum
computation.\cite{doucot:024505,milman:020503}

The compass model is a spin model on simple-cubic lattices in $d$ dimensions of size $N=L^d$ defined by the Hamiltonian
\begin{equation}
\label{eq:hamiltonian1}
\mathcal{H}=\sum_{i} \sum^d_{k} J_k S^k_i S^k_{i+e_k}\,,
\end{equation}
where $S^k_i$ represents the $k$-th component of a spin $\mathbf{S}$
at site $i$ and $i+e_k$ is the nearest neighbor of $i$ in the $k$
direction. In the classical case we have $\mathbf{S}\in O(d)$, or in a
more explicit vector representation with $\varphi$ and $\theta$ being
angles on the sphere, we use the expression
$\mathbf{S}^{\mathsf{T}}=\left(\cos(\varphi),\sin(\varphi)\right)$ and
$\mathbf{S}^{\mathsf{T}}=\left(\cos(\varphi)\sin(\theta),\sin(\varphi)\sin(\theta),\cos(\theta)\right)$
in two and three dimensions, respectively.  In the two-dimensional
(2D) quantum case $\mathbf{S}$ represents a spin-$1/2$ operator
$\mathbf{S}=(1/2)\,(\sigma_x,\sigma_z)$ and the Hamiltonian assumes the form
\begin{equation}
\label{eq:hamiltonian2}
\mathcal{H}=(1/4)\,\sum_{i}\left( J_x \sigma^x_i \sigma^x_{i+e_x} + J_z \sigma^z_i \sigma^z_{i+e_z}\right)\,,
\end{equation}
where we have chosen the $z$ instead of the $y$ direction as a matter
of convenience (usually we take $S^z$ as the quantization component
in quantum Monte Carlo). In this work the coupling constants
 are taken to be equal, $J_k=J$,  and positive although the sign plays no role
since it can be transformed away on bipartite lattices ($L$ must be even).

Recent contributions in the literature have explicitly investigated
the properties of the 2D compass model for both the classical and
quantum Hamiltonian. Analytical and Monte Carlo
work on the classical case proved the existence of
a directional-ordering transition at finite-temperatures and it was
argued that this transition belongs to the 2D Ising universality
class.\cite{mishra:207201} Using exact diagonalization techniques and Green-function Monte
Carlo the energy spectrum of low lying states was analyzed for the
quantum model in detail.\cite{doucot:024505,dorier:024448} These
studies provided the key result that the ground state is exponentially
degenerate possessing a degeneracy of $2\times2^L$. This turns the
relatively simple Hamiltonian into a hard problem comparable to
frustrated magnets. 
Later work\cite{chen:144401} determined the nature
of the quantum phase transition to be of first order when driving the
system by changing the coupling ratio $J_x/J_z$. A variant of the
model possessing a similar quantum phase transition was finally
analyzed in one dimension.\cite{brzezicki:134415} In a recent Letter
\cite{tanaka:256402} the finite temperature properties of the quantum
compass model were analyzed for the first time by means of a world
line quantum Monte Carlo scheme based on the Suzuki-Trotter
discretization. The authors conclude with the intriguing effect, that
the presence of random site dilution has much weaker effects on
criticality for quantum degrees of freedom than for classical ones.
The numerical analysis supporting this conclusion is, however, based
on rather small lattice sizes and the quality of the quantitative
results is modest and in view of our results reported below would
need further investigations.

Due to the relevance of the model and the potential implications for
future applications it would be desirable to have a more precise
understanding of the critical behavior at the directional-ordering
transition in the quantum compass model. The purpose of this work is
to tackle this problem with a comprehensive Monte Carlo study for both
the classical and quantum case where we will focus here on the
non-disordered case. Our motivation to restudy the classical case is
to gain as much experience as possible about the transition and
difficulties that may arise in the Monte Carlo sampling and data analysis.
Using this experience a large-scale simulation of the quantum compass
model in 2D will follow in the second part. The next section
introduces the methods and tools we used to accomplish this. Section
\ref{class2d} contains our results for the classical compass model and
Sec.~\ref{quantum2d} the respective analysis for the quantum case. We
close in Sec.~\ref{sec:summary} with a summary and our conclusions.

\section{\label{sec:methods}Observables and Methods}
\begin{figure}[b]
\begin{minipage}{0.49\columnwidth}
\includegraphics[bb=132 360 475 707, clip, width=0.95\textwidth]{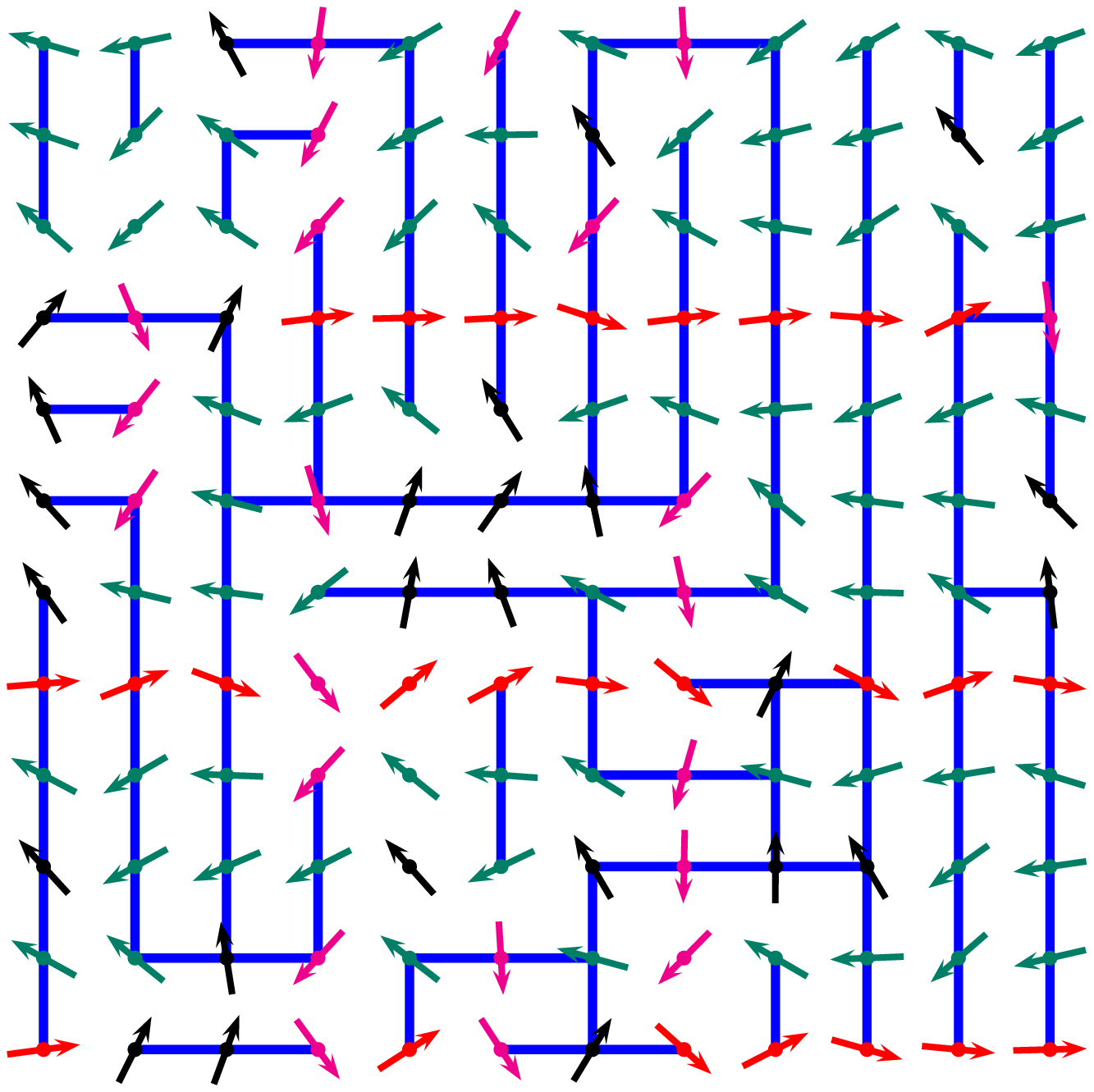}
\end{minipage}
\begin{minipage}{0.49\columnwidth}
\includegraphics[bb=132 360 475 707, clip, width=0.95\textwidth]{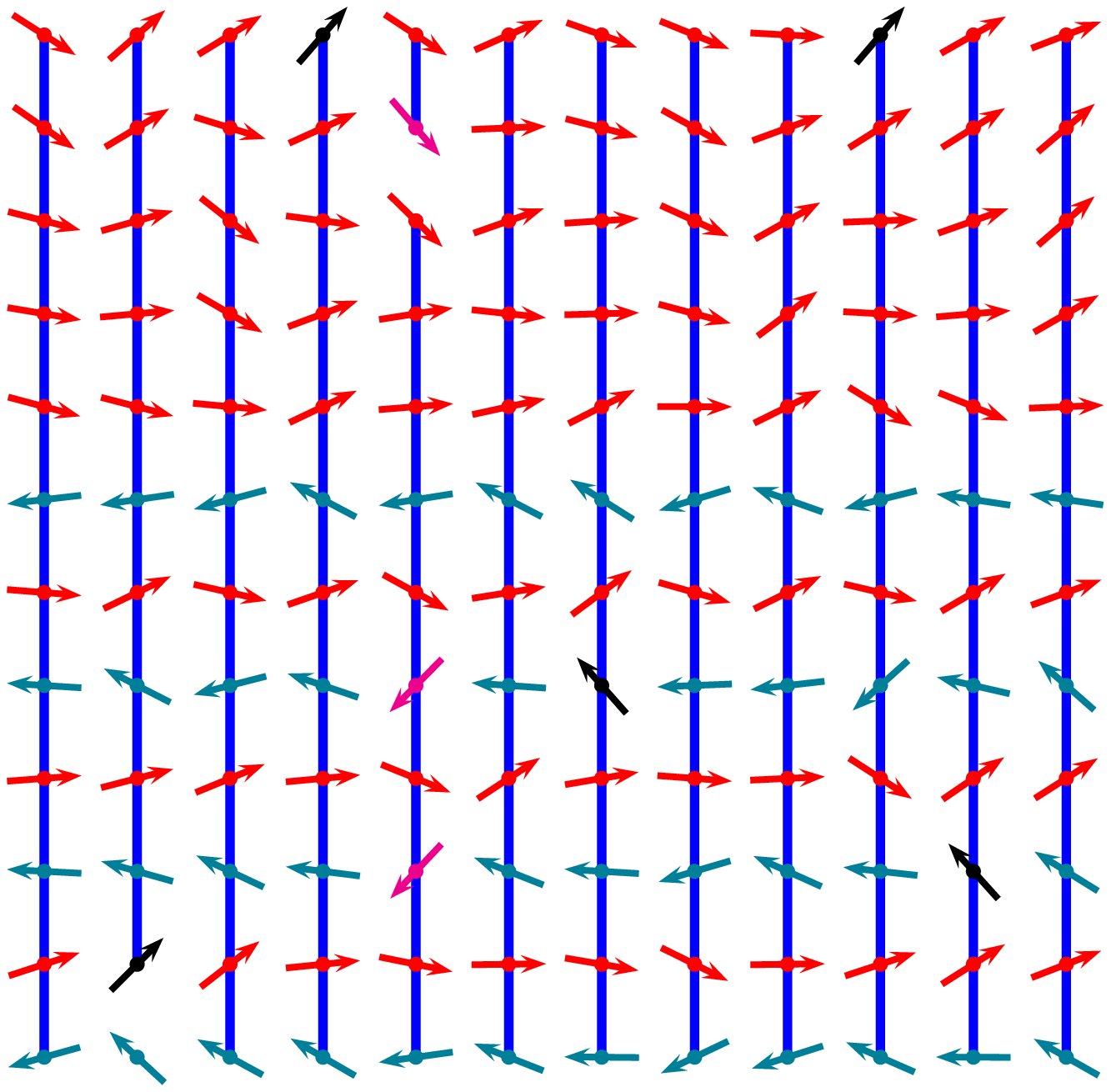}
\end{minipage}
\caption{\label{fig:disorderorder}(Color online) Visualization of
  different phases in the 2D compass model. \textit{Left:} For
  $T>\Tcr$ the system is disordered and the distribution of bonds
  possessing less than average bond energy (thick lines) is rather random.
  \textit{Right:} For $T<\Tcr$ the prevalent correlations order into
  one direction, i.e. the system is in a directionally-ordered state.
  The pictures are snapshots of a Monte Carlo simulation of the
  classical model with $L=12$ at $T=0.3J$ and $T=0.10J$ respectively
  (ferromagnetic representation). The small arrows indicate the spin
  degree of freedom.}
\end{figure}

\subsection{Observables}
 
In this section we describe the observables that are used to
characterize the phases and to probe the phase boundaries of the
compass model. The basic quantity is the total energy $E = H =\sum_k E_k$ and the corresponding heat capacity $C=\partial{E}/\partial
T$. With $E_k=J_k S^k_i S^k_{i+e_k}$ we denote the energy along the
$k$-th direction or on $k$-bonds in the system. Using this definition a
useful order parameter in 2D can
then be defined as\cite{mishra:207201,tanaka:256402}
\begin{align}
\label{eq:orderparameter}
D&=\frac{1}{N} \left| \sum_{i} \left( J_x S^x_i S^x_{i+e_x} - J_z S^z_i S^z_{i+e_z}\right) \right|\,,\\
\nonumber &=\frac{1}{N} \left| E_x - E_z \right|\,,
\end{align}
where $N=L^2$. The quantity
$D$ measures the excess energy in one direction compared
to the other direction. If $D>0$ the system is said to
possess long-ranged orbital or directional order whereas for
${D=0}$ the system is disordered.
An alternative definition for the order parameter 
\begin{equation}
D^\prime=\frac{1}{N} \left\langle \min{E_k} - \sum_{k=1}^{d=2} E_k / d \right\rangle\,,
\label{eq:ordergeneral}
\end{equation}
can be used to give a visualization and characterization of the
different phases as in Fig.~\ref{fig:disorderorder}. On the lattice we
thereby mark all bonds which have less than the average bond energy
(those that contribute most to the partition function) and look at the
global structure of the resulting bond clusters. In the disordered phase we expect
rather random clusters whereas the ordered phase is characterized by
clusters which are directionally ordered and independent of each other
(dimensional reduction). Note, that in two dimensions $D$ and
$D^\prime$ are actually the same quantity up to a constant
factor, because $D=2D^\prime$. However, Eq.~\eqref{eq:ordergeneral}
provides the general possibility to define an order parameter in any
dimension $d$, which might be useful for future studies.  In order to
investigate the universality class of the phase transition we further
look at the susceptibility $\chi$ and Binder parameter $Q_2$ which are
respectively defined as
\begin{equation}
\chi=N\left(\langle D^2 \rangle - \langle D \rangle^2 \right)\,,\quad Q_2=1-\frac{1}{3}\frac{ \langle D^4 \rangle}{\langle D^2 \rangle^2}\,,
\end{equation}
where $\langle D^n \rangle$ denotes an average of the $n$-th moment computed from the time series of $D$.

For the susceptibility we expect a finite-size scaling behavior of the form
\begin{equation}
\label{eqn:ffsgamma}
\chi\sim L^{\gamma/\nu}\,,
\end{equation}
at the critical point with $\nu$ being the correlation length
critical exponent and $\gamma$ the exponent for the susceptibility.
Neglecting corrections to scaling, the Binder parameters for different lattices sizes $L$ should ideally
cross at the critical temperature $T_\mathrm{c}$. In any case, the behavior of the
crossing points of $Q_2$ at lattice sizes $L$ and $2L$ should approach $T_\mathrm{c}$ like
$L^{-1/\nu-\omega}$ if we have corrections to scaling ($w<\infty$).

\subsection{Monte Carlo methods}
Ordinary Metropolis Monte Carlo simulations are used for the classical
model where we update each spin sequentially. During the
thermalization procedure we adjust the proposed moves such that
an average acceptance rate of about $50\%$ is obtained at each
temperature.  As it already becomes apparent from simulations on very
small lattice sizes $L$ that the system suffers from huge
autocorrelation times we add a parallel tempering (PT)
scheme\cite{geyerPT,hukushimaPT}, where we propose to exchange spin
configurations between simulation threads at different temperatures
$T_i$. This exchange is attempted every $n$ sweeps, where $n$ is
typically in the range $2$ to $20$.  By tracking individual
configurations we make sure all temperatures are seen and that
sufficient diffusion through temperature space is performed. For
simplicity the simplest PT scheme is used, meaning that an equidistant
temperature spacing between neighboring processes is chosen. As a
result a reduction of autocorrelation times by two orders of magnitude
is achieved which pays off in comparison to little longer simulation
times.

In case of quantum spin degrees of freedom, we employ a quantum Monte
Carlo (QMC) procedure based on the stochastic series expansion
(SSE)\cite{PhysRevB.43.5950} technique originally developed by Sandvik. Our own implementation is based on the
(directed) loop scheme\cite{PhysRevE.66.046701} supplemented by ideas of Ref.~\onlinecite{alet:036706}. Recall that the principle of SSE is sampling the
series expansion of the quantum partition function
\begin{align}
\label{eq:z}
\nonumber
Z&=\mathrm{tr}\left(\exp(-\beta \mathcal{H})\right)=\sum_\alpha \sum_n \frac{(-\beta)^n \langle \alpha| \mathcal{H}^n |\alpha\rangle}{n!}\,,\\
&=\sum_{b_i\in S_n} \sum_\alpha \sum_n \frac{\beta^n}{n!}\langle \alpha | \prod_i^n \mathcal{H}_{b_i} | \alpha \rangle\,.
\end{align}
by a Markov chain stochastic process, where $\beta=1/k_\mathrm{B}T$ is the inverse temperature. The last line of Eq.~\eqref{eq:z}
is the central starting point\cite{PhysRevB.43.5950} of the method because it
specifies the configuration space (and the weights) in which the
sampling takes place. A configuration lives in the product space of spin configurations
$|\alpha\rangle$ times the space of all possible sequences (or
permutations) $S_n$ of $n$ bond operators (or vertices) $\mathcal{H}_{b_i}$. The
degrees of freedom are thus $|\alpha\rangle$, $n$, and $S_n$, which
are sampled by the usual combination of diagonal, non-diagonal, and spin flip
updates.\cite{PhysRevE.66.046701}

In the case of the compass model the bond operators $\mathcal{H}_{b_i}$ can be derived from
the Hamiltonian \eqref{eq:hamiltonian2} as
\begin{equation*}
\mathcal{H}_{b}\in
\begin{cases}
S^z_i S^z_j & \text{if $b$ is a $z$ bond}
\\
\{S_i^+S_j^+,S_i^-S_j^-,S_i^+S_j^-,S_i^-S_j^+\} & \text{if  $b$ is a $x$ bond}\,,
\end{cases}
\end{equation*}
where the appearance of pure $S_i^{-}S_j^{-}$, and $S_i^{+}S_j^{+}$
terms are a notable difference to an ordinary Heisenberg model. Here
$S^+$ and $S^-$ refer to creation and annihilation operators and the
subscripts $i,j$ are the two sites of the bond $b$.  Simulations of
the quantum compass model are furthermore more involved since the
Hamiltonian dictates an asymmetry between bonds in $x$ and $z$
direction, allowing no spin flip operators of type $S^{\pm}S^{\pm}$ to
reside on $z$-bonds. On the other hand, there are {\it a priori} no
diagonal terms $S^zS^z$ on $x$-bonds. However, since non-diagonal
terms can only be introduced into the SSE configuration space after
the diagonal-update (non-diagonal operators must be present!) we are
therefore forced to introduce a positive non-zero energy shift
$\epsilon$ into the Hamiltonian of Eq. \eqref{eq:hamiltonian2}.  As a
consequence both non-diagonal and diagonal terms may reside on
$x$-bonds. On $z$-bonds only diagonal terms are allowed.
Unfortunately, this asymmetry in the operator representation cannot be
transformed away by a simple ``symmetrizing'' rotation of the
Hamiltonian because of emerging minus sign problems. Note finally that
the non-zero energy shift $\epsilon$ has an effect on the order
parameter $D$ since it influences the number and the distribution of
bond operators in the operator sequence.\footnote{This is an effect caused by defining the quantity $D$ as in Eq.~\eqref{eq:orderparameter} and by taking the absolute value $||$ explicitly from the time series of $E_x-E_y$.
A better way would be to take $D^\star=\sqrt{\langle (E_x-E_y)^2\rangle}$ which is not dependent on $\epsilon$. 
Then $D^\star$ is the true quantum estimate but is much harder to sample in SSE resulting in larger error bars.
$D$ is therefore an approximation to $D^\star$ which becomes almost perfect in the ordered phase. For our main objective of criticality this is, however, irrelevant and we choose a quantity which is more accurately to measure.}
This effect can cause
additional finite-size contributions also in the susceptibility and
the Binder parameter which vanish in the thermodynamic limit and for
$T\to 0$. We have checked that at $L=16$ no difference could be
detected in the susceptibility maxima locations for $\epsilon=0.1,
0.5, 0.9$ within error bars. Here we work with $\epsilon=0.5$ at all
lattices sizes.

Since simulations of the quantum model display the same rapid
critical slowing down as the classical model we perform additional
quantum PT updates\cite{PhysRevB.65.155113} in the same manner as described
above.
We implemented both the classical and quantum Monte Carlo PT
scheme on parallel architectures with the restriction of shared memory
access for fast communication between processes. This is essential
since PT updates are done rather often. 

For data analysis purposes we use well-known multi-histogram
techniques to optimally combine simulations at different temperatures.
Those techniques are available for both the
classical\cite{ferrenberg:multi} and quantum
cases\cite{PhysRevLett.90.120201,Troyer_multihist}. In combination
with optimization routines like the Brent method\cite{brent} they
allow rather systematic and unbiased estimation of pseudocritical
temperatures from peaks in the susceptibility.

\subsection{Boundary conditions}
\begin{figure}[b]
\begin{minipage}{0.2\textwidth}
\psfrag{l}{{\rm pbc}}
\includegraphics[width=\textwidth]{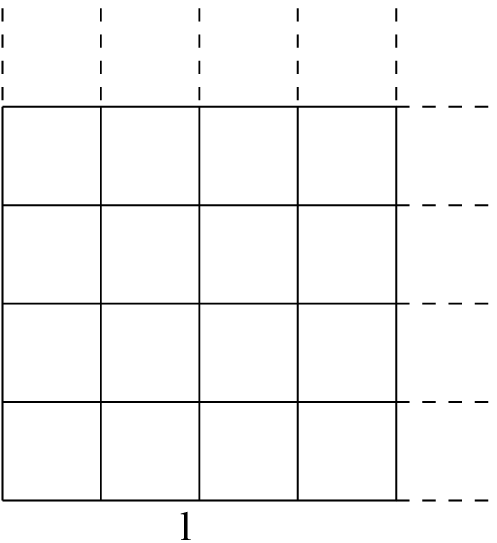}
\end{minipage}\makebox[0.5cm]{}
\begin{minipage}{0.2\textwidth}
\psfrag{l}{{\rm abc}}
\includegraphics[width=\textwidth]{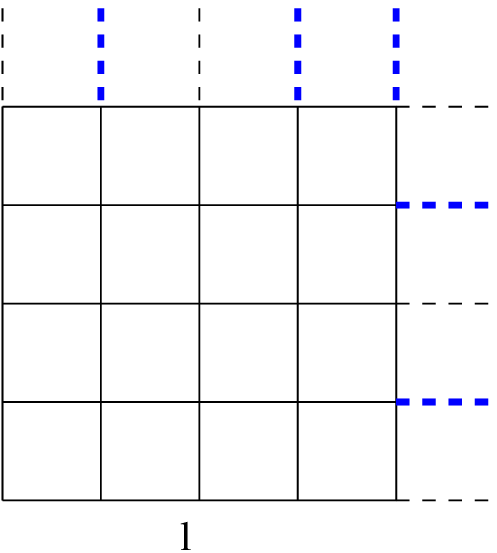}
\end{minipage}
\caption{\label{fig:bc}(Color online) Visualization of the different
  boundary conditions used in this work. {\it Left:} Ordinary
  periodic boundary conditions. All bonds carry the same coupling and
  the dashed bonds connect the spins across boundaries. The topology is a
  torus. We refer to this case as pbc. {\it Right:} So called
  ``annealed'' boundary conditions (abc).\cite{mishra:207201} Here
  the sign of the couplings on the dashed boundary bonds may fluctuate
  dynamically resulting in an additional degree of freedom.  As an
  example we draw some thick bonds indicating a negative coupling.}
\end{figure}
\begin{figure*}
\begin{minipage}{0.33\textwidth}
\includegraphics[width=\textwidth]{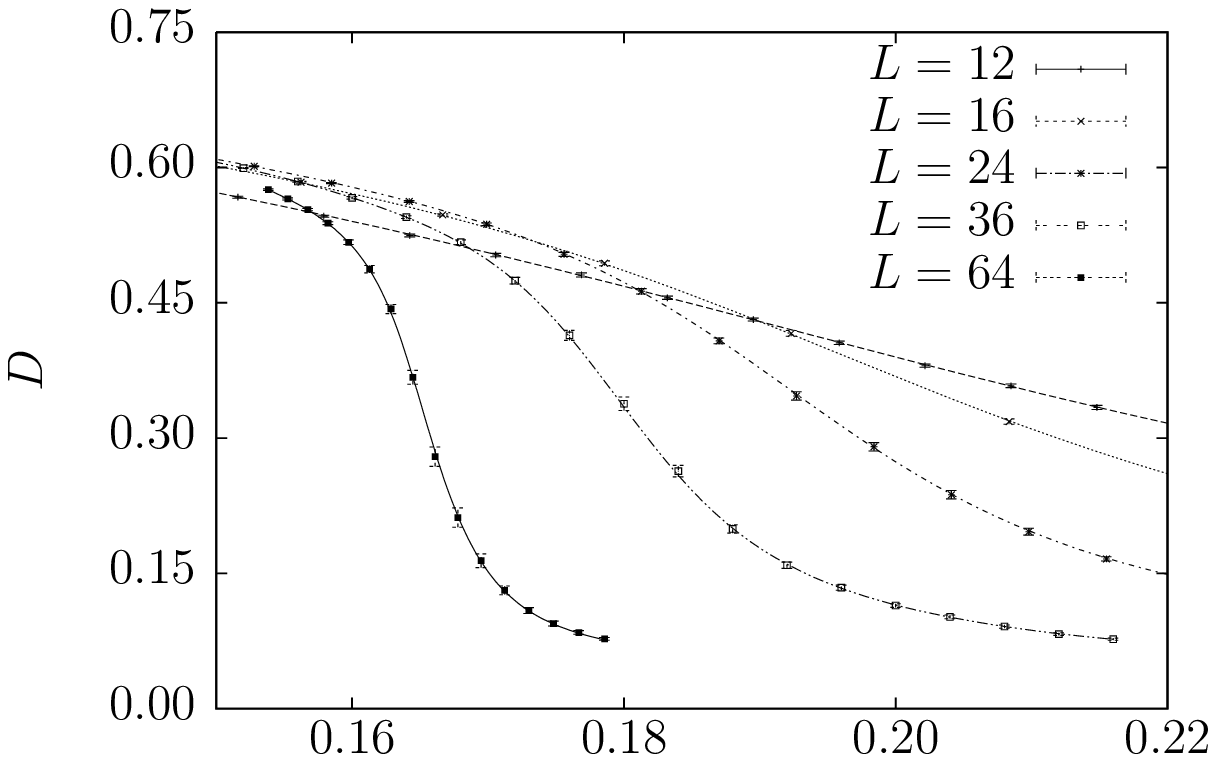}
\includegraphics[width=\textwidth]{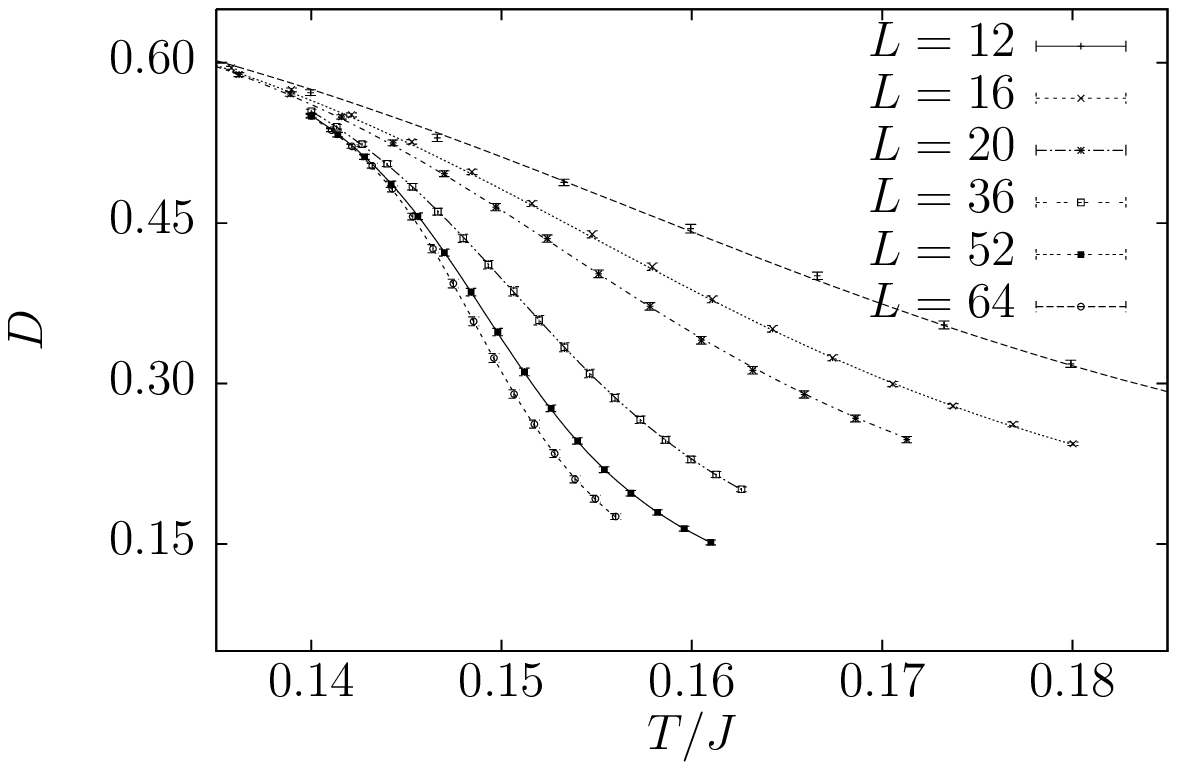}
\end{minipage}
\begin{minipage}{0.33\textwidth}
\includegraphics[width=\textwidth]{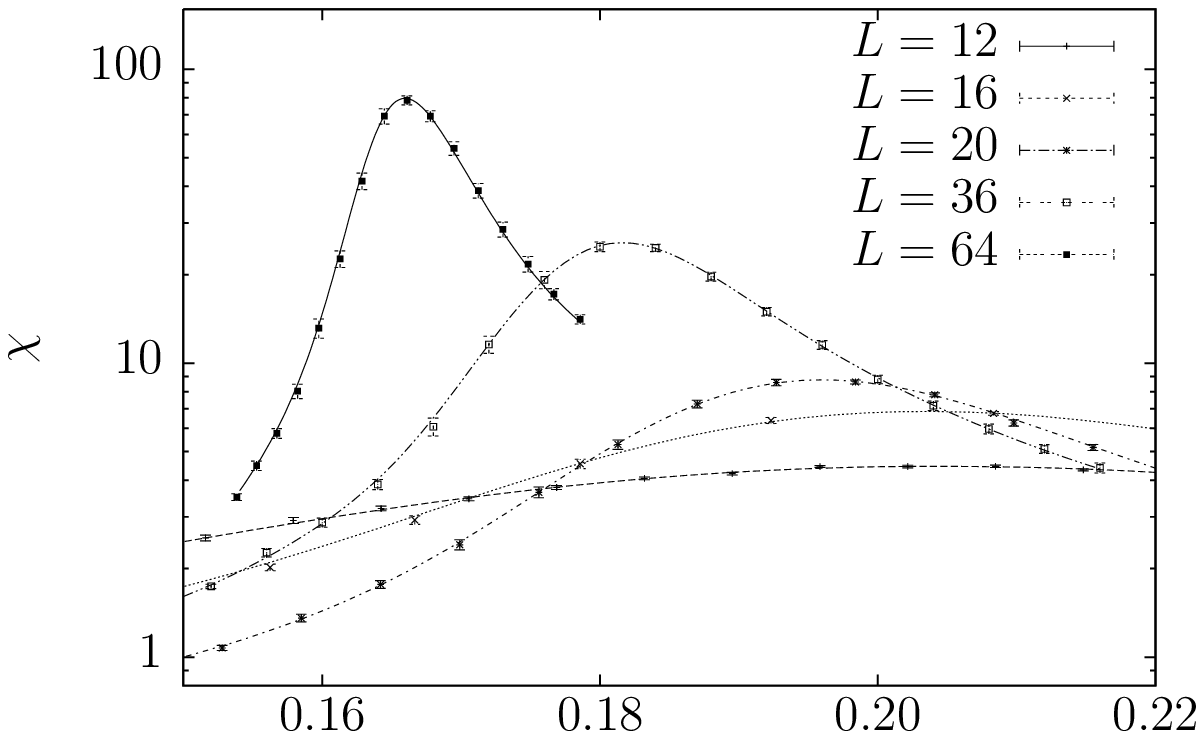}
\includegraphics[width=\textwidth]{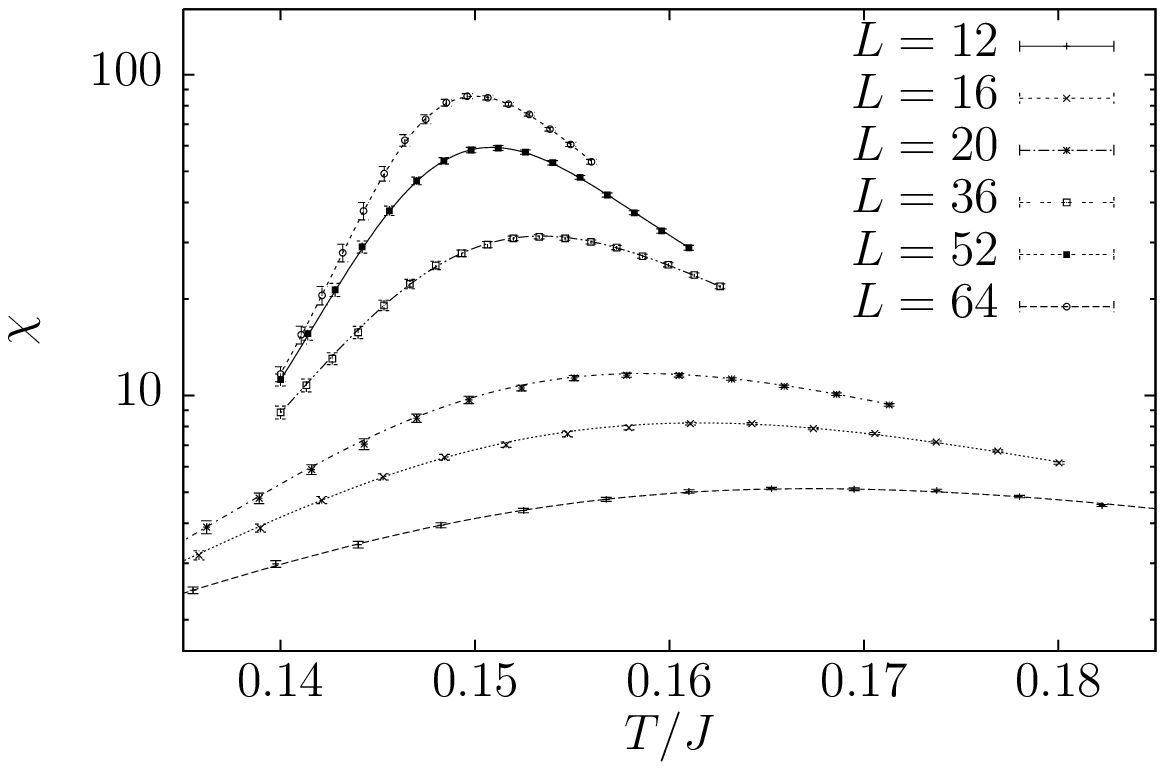}
\end{minipage}
\begin{minipage}{0.33\textwidth}
\includegraphics[width=\textwidth]{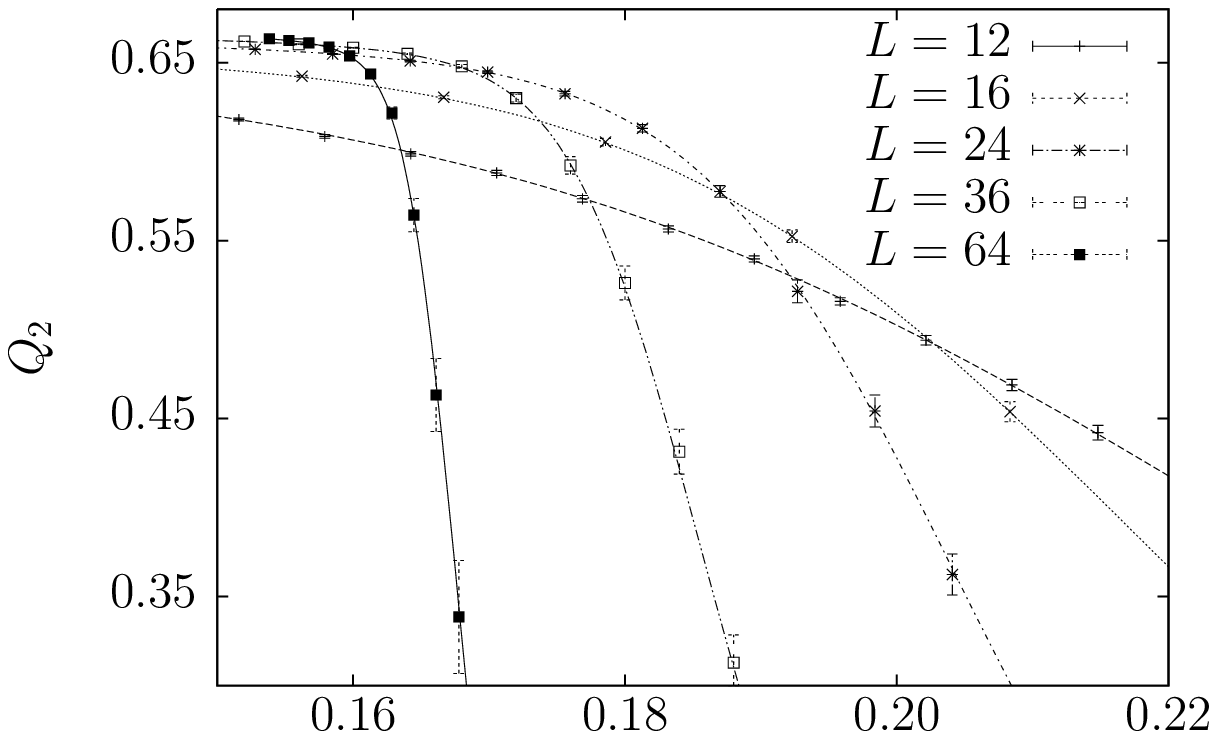}
\includegraphics[width=\textwidth]{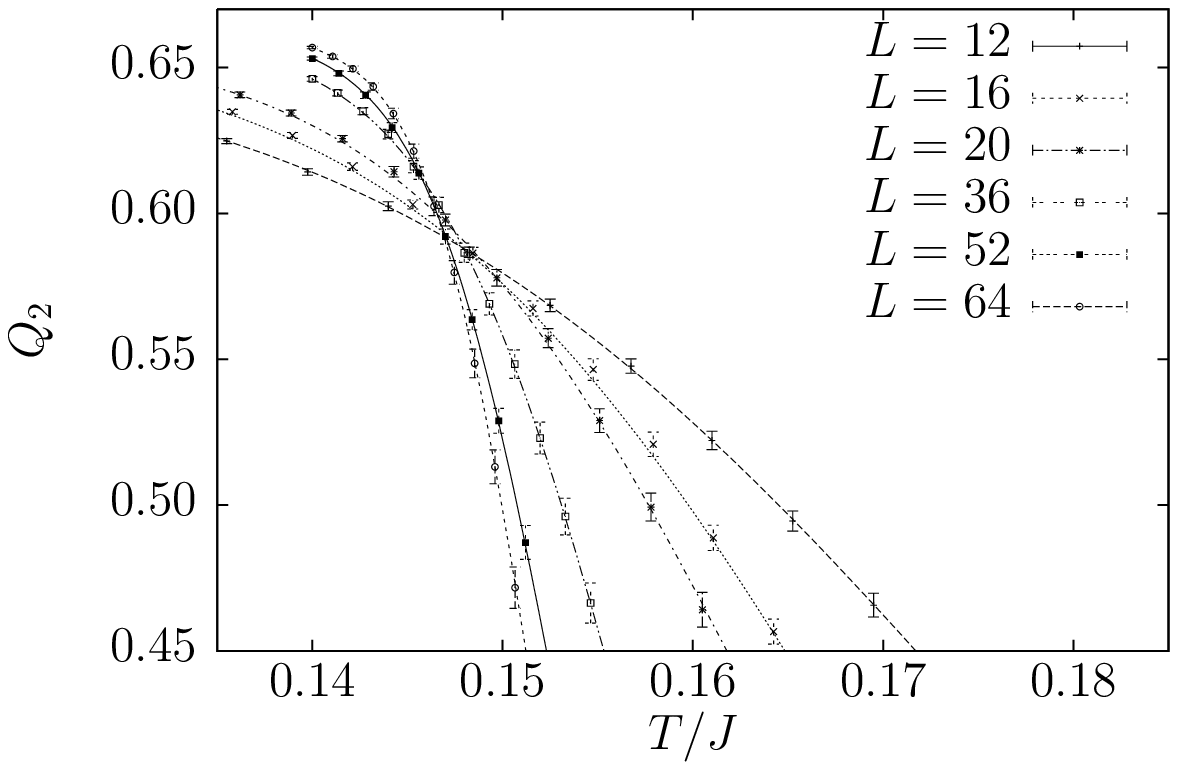}
\end{minipage}
\caption{\label{fig:result2d} Data for the 2D classical compass model
  obtained by Monte Carlo simulations. The top row displays the
  results for periodic boundary conditions and the bottom row for
  annealed boundary conditions. Note the the temperature ranges are
  different for both cases and that not all lattice sizes are shown
  for better readability. The lines through the data points are
  obtained from the multi-histogram analysis. {\it Left:} The order
  parameter $D$ as a function of different lattice sizes $L$. {\it
    Middle:} The susceptibility $\chi$ of the order parameter. {\it
    Right:} The Binder parameter $Q_2$.}
\end{figure*}

Ordinarily, the vast majority of Monte Carlo simulations are performed
using periodic boundary conditions (pbc) which map the lattice onto a
torus topology using the assumption that free-energy contributions
from the surface are thereby minimized.  In contrast to this approach,
Mishra {\em et al.}\cite{mishra:207201} argue in their recent
contribution that periodic boundary conditions might not be optimal in
the case of the compass model. Instead, they introduce special, so
called \textit{annealed} boundary condition (abc) to arrive at their
Monte Carlo results.  Since a detailed comparison between these two
boundary conditions has to our knowledge not been done, we will
explicitly study and compare their effect on the finite-size scaling
behavior for the classical compass model. This comparison is
especially interesting in view of the fact that we may not easily
apply the annealed case to quantum Monte Carlo since it induces a
minus sign-problem. A characterization and understanding of the
scaling behavior for periodic boundary conditions would therefore be
of advantage before studying the quantum case.

Figure \ref{fig:bc} displays these two types of boundary conditions as
a sketch. The topology of the annealed boundary condition is the same
as for periodic boundary conditions. The annealed case is special
because the sign of couplings on bonds  across the border
may fluctuate dynamically according to the Boltzmann distribution.
The bond sign is therefore an additional degree of freedom in the
Monte Carlo update rendering the simulations somewhat more complex.
\section{\label{class2d}The classical compass model in 2D}
In this section we start the presentation of our simulation results.
We consider firstly Monte Carlo simulations of the 2D classical
compass model. The main purpose of this section is to give an explicit
comparison between the different boundary conditions introduced in the
last section. To this end we run simulations for both cases and
compare the observables of Sec.~\ref{sec:methods} and their
finite-size behavior. Figure~\ref{fig:result2d} gives an overview of
our Monte Carlo estimates for $D$, $\chi$ and $Q_2$. There, the top
row contains results for periodic boundary conditions and the bottom
row for annealed boundary conditions. Knowing the different behavior
of reaching the thermodynamic limit is useful in order to appreciate
results of our simulations which follow in subsequent sections.

Simulations are done using lattice sizes {$L
  =\{10,12,16,24,32,36,48,64,128\}$} periodic bc and
${L=\{10,12,16,20,24,36,52,64\}}$ for annealed bc, typically taking
about $10^5$ measurements per data point after an equilibration phase
of $10^4$ sweeps. By the behavior of the order parameter in
Fig.~\ref{fig:result2d} (left) it is immediately evident that there is
a phase transition and that directional order with $D>0$ is realized
at low temperatures. We secondly observe that the order parameter for
the pure periodic case has a slow convergence for small lattice sizes
while for larger sizes it suddenly moves considerably. In contrast,
the data for the annealed case show a much smoother movement towards
the infinite-volume limit and it is evident that finite-size effects
are drastically reduced.  A difference like this is actually expected
for different boundary conditions. The crucial and interesting
question is whether the two boundary conditions lead to the same
critical temperature in the infinite volume limit where boundary
effects should vanish.

We therefore obtain an estimate of the critical point $T_\mathrm{c}$
in the thermodynamic limit by fitting the pseudocritical temperatures
$T_\mathrm{c}(L)$ taken from the peaks of the susceptibilities in
Fig.~\ref{fig:result2d} (middle) at lattice size $L$ to the
finite-size scaling ansatz
\begin{equation}
  T_\mathrm{c}(L)=T_\mathrm{c} + bL^{-1/\nu}(1+cL^{-\omega})\,.
\label{eqn:ffs}
\end{equation}
Here $b,c$ are some constants and $\omega$ is an exponent
describing corrections to scaling. In a first step, we assume nothing
about the value for the correlation length exponent $\nu$ and leave it
as fit parameter.
\begin{figure}
\includegraphics[width=\columnwidth]{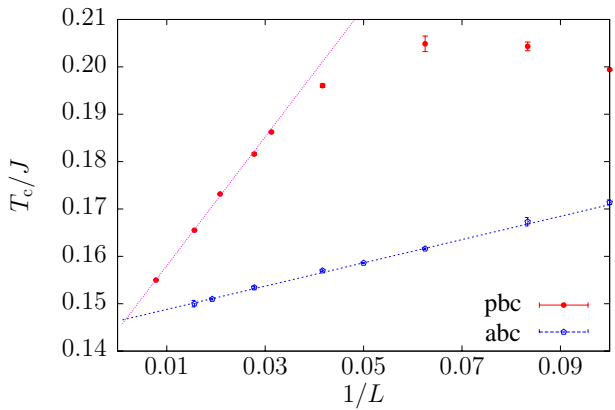}
\caption{\label{fig:ffsorder_cmp}(Color online) Determination of the
  critical temperature $T_\mathrm{c}$ from finite-size scaling of the
  pseudocritical temperatures determined from susceptibility peaks.
  The two curves correspond to different boundary conditions which
  trigger a completely different convergence to the critical point.
  The lower curve is obtained for annealed boundary conditions (abc) and
  shows the superior scaling compared to periodic boundary conditions (pbc).
  Lines are fits to Eq.~\eqref{eqn:ffs} neglecting the correction term
  $\omega$.}
\end{figure}
The fitting procedure to the data in Fig.~\ref{fig:ffsorder_cmp}
yields $T_\mathrm{c}=0.144(2)J$ from periodic boundary conditions and
$T_\mathrm{c}=0.1461(8)J$ from annealed boundary conditions and the
estimate for the correlation length critical exponent is $\nu=0.98(4)$
which we take from the straight line fit for the annealed case. Both
results agree within error bars. The annealed value yields a much more
accurate estimate since here the asymptotic scaling regime sets in
much earlier and we have more points available for fitting.  These
numerical estimates are in accordance with the value
$T_\mathrm{c}=0.147(1)J$ obtained in Ref.~\onlinecite{mishra:207201}.
With our value for $\nu$ we support the claim that the transition is
of 2D Ising type. To further confirm this conjecture we also determine
the exponent $\gamma$ associated with the susceptibility $\chi$. For
lattice sizes large enough ($L>20$) we obtain $\gamma/\nu=1.73(4)$ from the annealed case
(see Table~\ref{tab:results} and Fig.~\ref{fig:gamma} below) which is
again consistent with 2D Ising universality.  In a second step, we can
now assume Ising universality to be given to improve the fit.  Using
$\nu=1$ as a fixed parameter the improved value for critical
temperature is $T_\mathrm{c}=0.1464(2)J$.

Let us now turn to a discussion of the Binder parameter $Q_2$
displayed in Fig.~\ref{fig:result2d} (right). For the annealed case a
nice crossing of curves at the critical temperature can be observed
and our estimate for the Binder parameter at the crossing point
(taking the three largest lattice sizes) is $Q_2=0.61(1)$. This is
roughly the known value for the 2D Ising model, which -- however -- is
usually obtained for {\it periodic} boundary conditions.\cite{Kamieniarz} Using the
observed crossing behavior, the Binder parameter supplies a natural
third check of the critical temperature and the critical exponent. We
hence apply our recently developed data collapsing
tool\cite{wenzel:nuclphysB2007} and obtain $T_\mathrm{c}=0.1465(4)J$
and $\nu=1.01(4)$ from the best data collapse. These values are again
fully consistent with our results above and give further confidence to
our analysis.
In contrast, the nice properties of the Binder parameter do not show
up for periodic boundary conditions, where it is hard to judge whether
curves for different lattice sizes cross in a single point at all.
Rather, we see strong finite-size effects and that the crossing points
for large lattice sizes move close to $2/3$, which is totally in
contrast to the expected behavior. It is known that different
boundary conditions cause a discrepancy (see for instance
Refs.~\onlinecite{Kamieniarz} and \onlinecite{selke-2006-51}) in the Binder crossings but such a
drastic behavior was unexpected.

In summary, our investigation for the classical model clearly show
that annealed boundary conditions are favorable because they
drastically reduce finite-size effects and yield good scaling
properties for the finite-size analysis. With periodic boundary
conditions much larger lattice sizes need to be investigated in order
to obtain the critical temperature and to approach the right asymptotic scaling
regime. Additionally, our analysis shows that the Binder parameter for
periodic boundary conditions does not cross at the usual expected
value and that we may not use the crossing point (height) as a good
indication for the critical point, whereas for annealed conditions we
get good properties.  These effects are currently not properly
understood. By referring to the typical spin configuration in
Fig.~\ref{fig:disorderorder} it is, however, tempting to argue that
the dominant energy correlations (blue lines) wrap around the torus in
the ordered phase thereby forming some kind of closed loop
excitations.  These excitation appear to be more stable against
thermal fluctuations than open excitation. Annealed boundary
conditions seem to prohibit the formation of such loops leading to a
better scaling behavior.

\section{\label{quantum2d}The quantum compass model in 2D}
\begin{figure}
\begin{minipage}{\columnwidth}
\includegraphics[width=\textwidth]{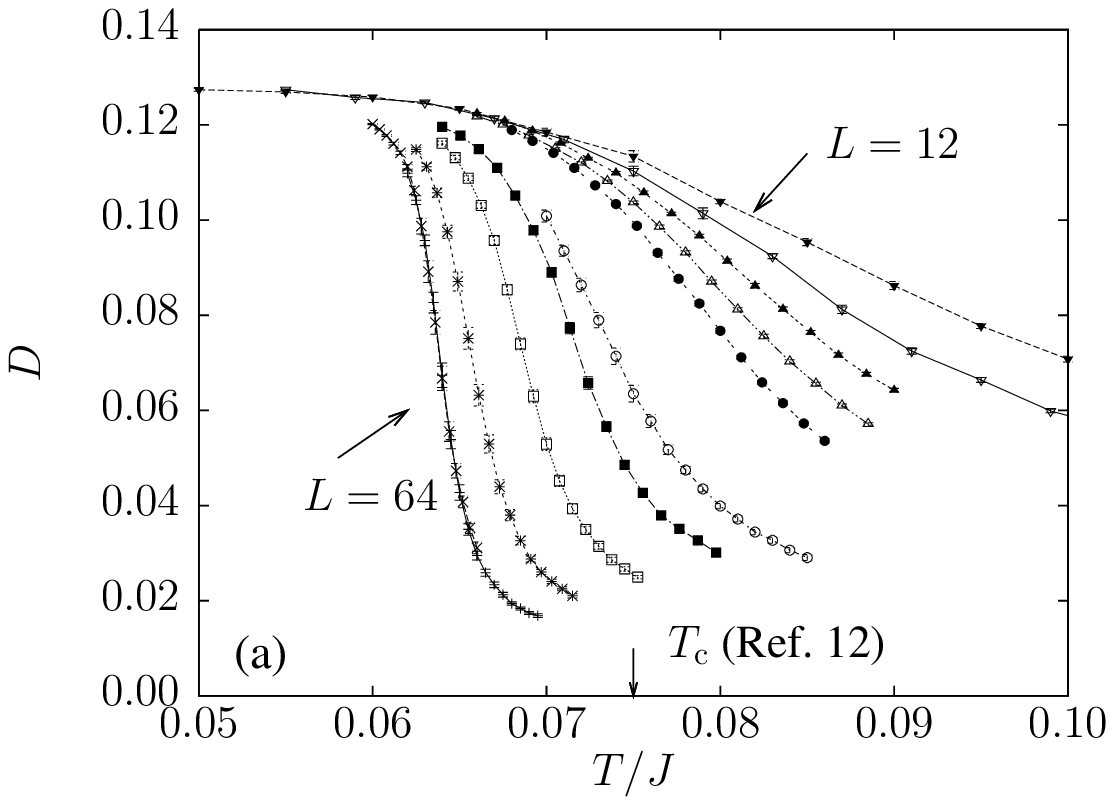}
\includegraphics[width=\textwidth]{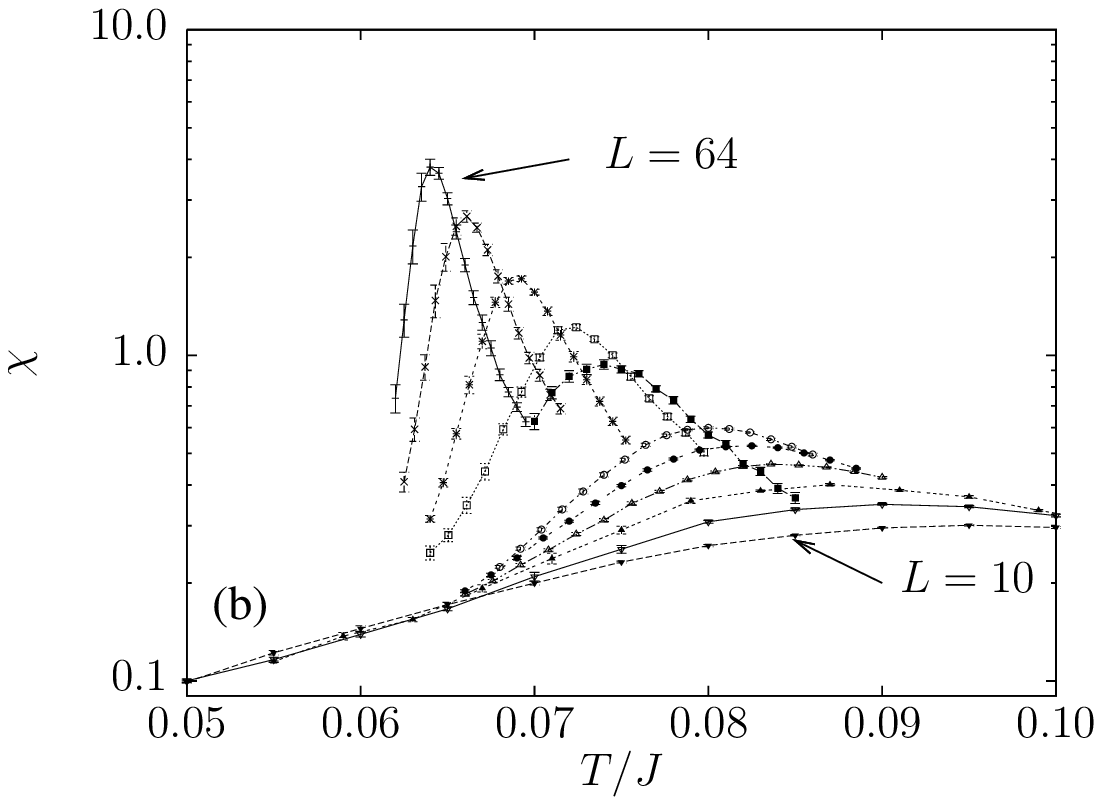}
\includegraphics[width=\textwidth]{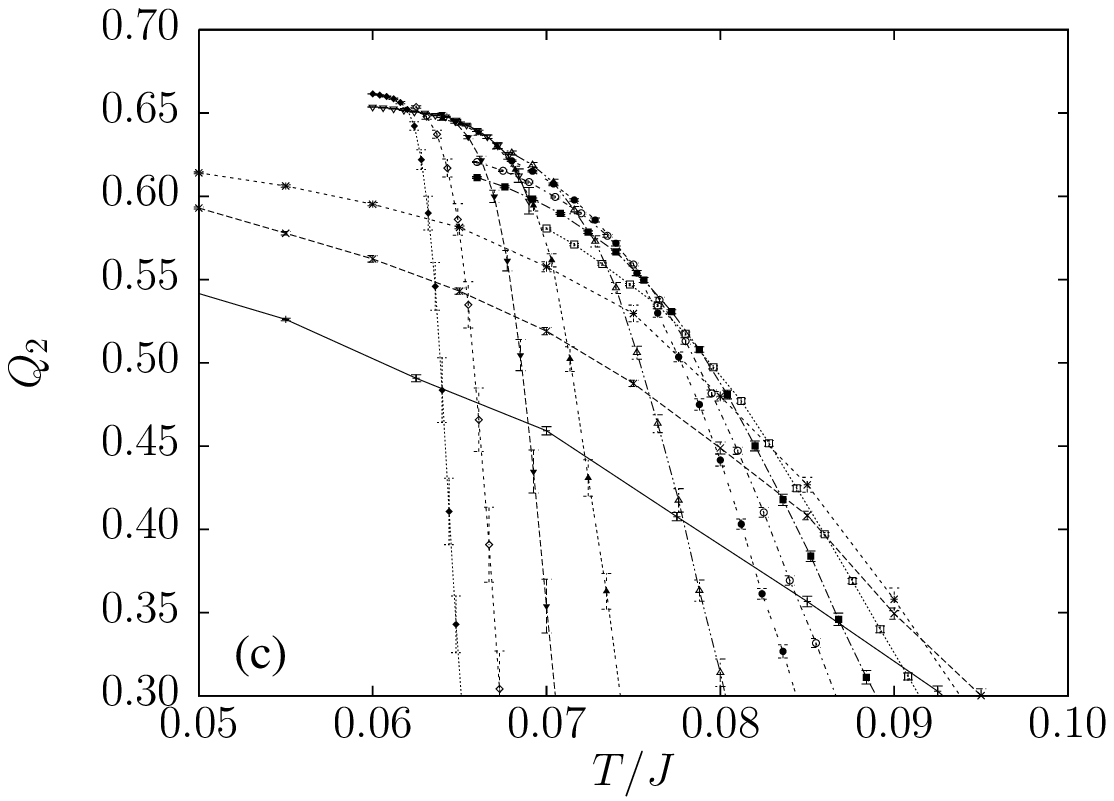}
\end{minipage}
\caption{\label{fig:qcm_orderandsus} QMC results for the 2D quantum
  compass model with periodic boundary conditions. All lines are a
  guide to the eye.  (a) The order parameter $D$ for lattice sizes
  ${L=\{12,14,16,18,20,28,32,40,52,64\}}$ displays a clear signal of a
  stable ordered phase at low temperatures. The arrow marks the
  transition temperature $\Tcr$ from Ref.~\onlinecite{tanaka:256402}.
  Our own data indicates a smaller value. (b) The susceptibility
  $\chi$ on a logarithmic scale for lattice sizes
  $L=\{10,12,14,16,18,20,28,32,40,52,64\}$. (c) the Binder parameter
  $Q_2$ in the quantum compass model with periodic boundary
  conditions, where steeper slope corresponds to larger $L$
  (neglecting $L=28$ and $L=48$ for better clarity). The qualitative
  behavior is the same as for the Binder parameter with periodic
  boundary conditions in the classical model. No common crossing point
  is present for the lattice sizes of this work.}
\end{figure}
Using the knowledge gained from simulations of the classical compass
model we turn to the discussion of the simulation results of the
quantum version.  Simulations are done using the stochastic series
expansion as outlined in Sec.~\ref{sec:methods}. The reader is
reminded that annealed boundary conditions, where the sign on boundary
bonds fluctuates, are not possible because such fluctuations induce a
sign problem in the quantum Monte Carlo scheme.

We therefore choose to simulate with periodic boundary conditions and
expect from Fig.~\ref{fig:ffsorder_cmp} that large lattice sizes might
be needed to see the right scaling and to obtain the infinite-volume
critical temperature. Using the parallel tempering scheme and the
reduction of autocorrelation times by two orders of magnitude, we were
finally able to simulate lattice sizes
${L=\{8,10,12,14,16,18,20,24,28,32,40,48,52,64\}}$, where the largest
one is about the limit one can reach in quantum Monte Carlo in
feasible time and resources at the moment.\footnote{In this
  temperature regime.} Our largest system size is about three times as
large compared to the simulations of Ref.~\onlinecite{tanaka:256402}.
A detailed check and verification of our
algorithm was done with data from full exact diagonalization (ED) on a
$4\times 4$ lattice. We use our own ED program with some implemented
symmetries\cite{dorier:024448} to reduce the dimension of the Hilbert
space, as well as the ALPS package\cite{ALPS} for smaller system
sizes. During the Monte Carlo runs, a total number of about $4\times
10^5$ measurements are typically taken after each sweep and $2\times
10^4$ sweeps are used for thermalization. Those numbers are, of
course, only meaningful with the additional information that we
construct as many loops in the non-diagonal update such that on
average $2n$ vertices are visited in the SSE configuration.
Figure~\ref{fig:qcm_orderandsus} shows the result for the order
parameter, the susceptibility and the Binder parameter obtained from the simulations in this
manner. The behavior of the order parameter shows a clear signal of a
transition from a disordered to an ordered state at small temperature,
evidently becoming more pronounced with increasing lattice size. This
proves the existence of a directional-ordering transition also in the
quantum case. In the ordered phase, the order parameter seems to take
on a value which is quite different from the classical case and the
order is furthermore less stable against thermal fluctuations as the
temperature regime in the quantum case is evidently much smaller.
Note that the overall estimate for $D$ also agrees roughly with data of
Ref.~\onlinecite{tanaka:256402}.

The dependence of the data on the lattice sizes is, as expected, qualitatively
similar to the classical case, i.e.,  the order parameter curves and
the susceptibility peaks shift considerably to lower temperatures for
larger lattice sizes.  This shift is in fact so large that it is
already obvious from Fig.~\ref{fig:qcm_orderandsus} (a) that the previous estimate of the
critical temperature in the literature is much to large.
\begin{figure}
\includegraphics[width=\columnwidth]{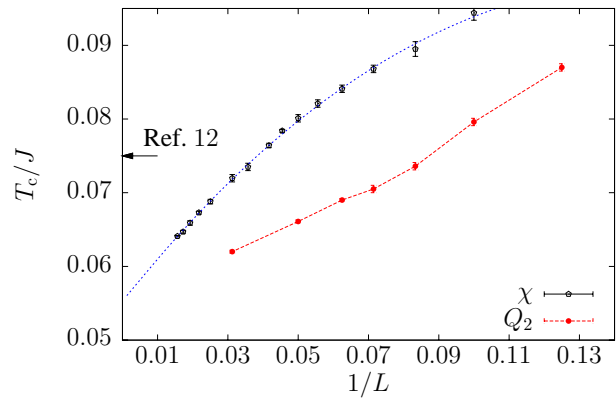}
\caption{\label{fig:scaling}(Color online) Finite-size scaling of pseudocritical temperatures for different
  lattice sizes obtained from the susceptibility. For small lattice sizes corrections to scaling are
  evident. For large lattice sizes 2D Ising scaling is reached
  yielding our estimate for the critical temperature of
  $\Tcr/J=0.055(1)J$. The curve trough the points represents a fit to
  Eq.~\eqref{eqn:ffs}. We also show the crossing points of
  the Binder parameters at $L$ and $2L$ for a consistency check. The arrow
  indicates the previous result of Ref.~\onlinecite{tanaka:256402}.}
\end{figure}
\begin{figure}
\includegraphics[width=\columnwidth]{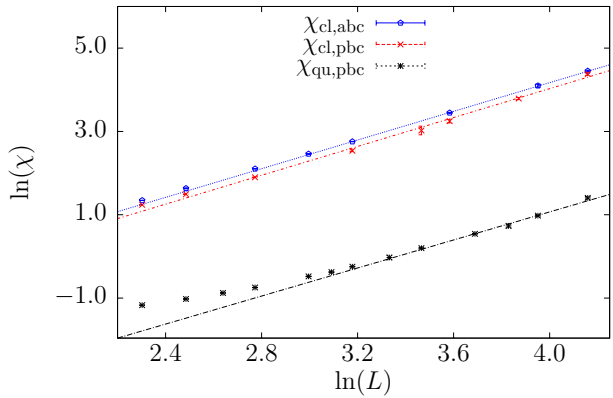}
\caption{\label{fig:gamma}(Color online) Plot of the susceptibility $\chi$ (at peak locations) versus system size $L$ for
  all different simulations in this work (classical annealead bc, classical periodic bc, quantum periodic bc in this order from top to bottom) on a double logarithmic axis. 
  The straight lines are fits to Eq.~\eqref{eqn:ffsgamma}. All cases are
  consistent with a value of $\gamma/\nu=1.75$.}
\end{figure}
Before we quantify this discrepancy for the critical temperature, we
draw our attention to the susceptibility and the Binder parameter in
Figs.~\ref{fig:qcm_orderandsus}(b),(c), both showing a behavior similar
to the classical case with periodic bc. We note especially that the
Binder parameter is again behaving oddly and that there is not a well
defined crossing seen at all at the lattice sizes simulated. A
crossing point might still be achieved for very large lattice lengths
$L$ but is certainly difficult to quantify since the value of $Q$ at
the crossing point is very close to $2/3$. Due to this observation the
Binder parameter is clearly not suited to determine the critical
temperatures by looking at the Binder crossings for small lattice
sizes, where the true behavior is just not seen.

Let us now determine an improved value for the critical temperature
with finite-size scaling from the maxima of the susceptibilities. As
in the classical case, we fit the pseudocritical values to the scaling
ansatz given in Eq.~\eqref{eqn:ffs}. To use as many data points as
possible we include corrections to scaling, described by the exponent
$\omega$, into the fit and leave all fit parameters free. Including
all lattice sizes we obtain $\Tcr=0.055(6)J$ and $\nu=0.9(2)$ with a
fit quality of $\chi^2/\mathrm{d.o.f}=0.66$. Those values, however,
are stable also for fitting windows starting at larger lattice sizes.
The precision for the critical exponent $\nu$ is rather low but agrees
with our expectation of 2D Ising universality within the error bar.
Under this assumption, we fix $\nu=1$ and repeat the fit procedure
yielding an improved estimate for the critical temperature as
$\Tcr=0.055(1)J$. The relative discrepancy with the previous estimate
of Ref.~\onlinecite{tanaka:256402} is approximately $36\%$.  As a
cross check for our analysis we further look at the scaling of the
crossing points of $Q_2$ at lattice sizes $L$ and $2L$, which is also
indicated in Fig.~\ref{fig:scaling}. We observe that this scaling is
consistent with the previous value from the susceptibilities but we do
not attempt a detailed fit by lack of enough data points. It
  is then also useful to obtain an independent estimate of the critical
  temperature from the maxima in the heat capacity $C$, which again gives
  consistent results but does not reach the accuracy of our previous
  analysis since $C$ is generally hard to sample in QMC at low temperatures.

To finalize our analysis, we determine the critical exponent $\gamma$
from the susceptibility of the order parameter $D$. For large lattice
sizes we expect a scaling according to Eq.~\ref{eqn:ffsgamma} which
can be tested by plotting $\ln(\chi)$ versus $\ln(L)$. This is done in
Fig.~\ref{fig:gamma} together with the data for the classical cases.
It is evident that asymptotic scaling sets in only for the largest
lattice sizes from which we obtain a value of $\gamma/\nu=1.68(8)$
consistent with 2D Ising universality, but not precise enough to be
absolutely conclusive.

\section{\label{sec:summary}Summary and Conclusions}
\renewcommand{\arraystretch}{1.2}
\begin{table}[t]
  \caption{\label{tab:results}Results for the critical temperature and critical exponents as obtained in this work.  The upper section contains the results for the classical model taken from annealed bc, which are all pairwise consistent. 
    The middle section summarizes our estimates for
    the quantum compass model for the cases with and without the assumption of 2D Ising behaviour ($\nu=1$). Both cases are consistent with each other.
    Lastly the values for $\gamma/\nu$ are summarized as obtained from the largest lattice sizes for the different simulation runs.}
\begin{tabularx}{\columnwidth}{Xllll}
\hline\hline 
 & $T_\mathrm{c}/J$ & $\nu$ & $\omega$ & $\chi^2/\mathrm{d.o.f}$ \\
\hline
no assumption & $0.1461(8)$ & $0.98(4)$ & $-$ & $1.3$ \\
2D Ising      & $0.1464(2)$ & $-$   & $-$ & $1.15$ \\
collapse $Q_2$ & $0.1465(4)$ & $1.01(4)$ & $-$ & $-$ \\
\hline
no assumption & $0.055(6)$  & $0.9(2)$ & $0.5(4)$ & $0.66$ \\
2D Ising      & $0.055(1)$  & $-$   & $0.8(2)$ & $0.61$ \\
\hline\hline
 & class. (abc) & class. (pbc) & quant. (pbc) & \\
\hline\
$\gamma/\nu$ &  $1.73(4)$ & $1.72(5)$ & $1.68(8)$ & \\
\hline\hline
\end{tabularx}
\end{table}
In this paper we reported on comprehensive Monte Carlo simulations of
the classical and quantum compass model. By comparing different
boundary conditions for the classical case, we showed that for
ordinary periodic boundary conditions one needs to go to very large
lattice sizes to see the right scaling and to get good convergence to
the critical point. In order to simulate large lattice sizes, we
implemented a parallel tempering scheme to counteract huge
autocorrelation times. Our results, which are summarized in Table
\ref{tab:results} are perfectly consistent with previous studies in
the literature for the classical model.  For the quantum model our
simulations are quantitatively at odds with earlier studies and we
provide here a new estimate for the critical temperature $\Tcr$. We
argued that this discrepancy might be explained by the huge
finite-size corrections originating from stable loop excitations
formed by correlation orderings which appear on the torus topology at
periodic boundary conditions. It appears that those excitations even
destroy the usual properties of Binder parameters.  Our analysis,
however, shows that one can still arrive at an estimate for
$T_\mathrm{c}$ at periodic boundary conditions provided that one takes
this effect into account. All critical exponents obtained in this
study give further support to the claim that 2D Ising universality
describes the directional-ordering transition in the 2D compass model.

Our findings for the quantum model might have an impact on the
conclusions of Ref.~\onlinecite{tanaka:256402} because a precise
estimate of $T_\mathrm{c}$ enters into the analysis of dilution
effects on the model. It is conceivable that the conclusion obtained
there are still qualitatively valid. For a true quantification
of the dilution effect, however, there is no way around performing a more
detailed investigation of larger lattice sizes. The knowledge gained in this
work should help to start such a study.

The precision of our results for the quantum model are still rather low
compared to many other systems of statistical physics. In this respect
it would be an interesting future project to devise and analyze
special boundary conditions for the quantum model with improved
finite-size scaling behavior compared to periodic boundary
conditions.

\acknowledgments We thank L.-H.~Tang and T.~Platini for discussions, and S.~Wessel for further suggestions.
S.W. acknowledges a PhD fellowship from the Studien\-stiftung des
deutschen Volkes, the kind hospitality of the statistical physics
group at the university Henry Poincare in Nancy and support from the
Deutsch-Franz\"osische Hochschule (DFH) under Contract No.
CDFA-02-07 as well as from the graduate school ``BuildMoNa.'' We also profited from a DAAD PPP programme with China.
Part of the simulations were performed on the JUMP facility of NIC at
Forschungszentrum J\"ulich under Project No. HLZ12.

\bibliographystyle{apsrev}
\bibliography{literature_nourl}

\end{document}